\documentclass[aps,twocolumn,superscriptaddress]{revtex4}

\usepackage{graphicx}
\usepackage{color} 
\usepackage{amsmath,amsthm,amsfonts,amssymb,bm}
\usepackage{times}
\usepackage{bbold}
\usepackage{epsf}
\usepackage{wasysym}
\usepackage[T1]{fontenc}
\usepackage[utf8]{inputenc}
\usepackage{ulem}

\newcommand{\half}{\mbox{$\textstyle \frac{1}{2}$}}
\newcommand{\ket}[1]{\left| #1 \right\rangle}
\newcommand{\bra}[1]{\left\langle #1 \right|}
\newcommand{\proj}[1]{\ket{#1}\bra{#1}}

\begin{document}

\title{Extracting Information from Qubit-Environment Correlations}
\author {John H. Reina}
\altaffiliation{E-mail: \texttt{john.reina@correounivalle.edu.co}}
\affiliation{Departamento de F\'isica, Universidad del Valle, A.A.
25360, Cali, Colombia}
\affiliation{Centre for Bioinformatics and Photonics---CIBioFI, Calle 13 No. 100-00, Edificio 320, No. 1069, Cali, Colombia}
\author {Cristian E. Susa}
\affiliation{Departamento de F\'isica, Universidad del Valle, A.A.
25360, Cali, Colombia}
\affiliation{Centre for Bioinformatics and Photonics---CIBioFI, Calle 13 No. 100-00, Edificio 320, No. 1069, Cali, Colombia}
\author {Felipe F. Fanchini}
\affiliation{Departamento de F\'isica, Faculdade de Ci\^encias, UNESP, Bauru, SP, CEP 17033-360, Brazil}

\date{\today}

\begin{abstract}

\bf{Most works on open quantum systems generally focus on the reduced physical 
system by tracing out the environment degrees of freedom. Here we show that the 
qubit distributions with the environment are essential for a thorough analysis, and 
demonstrate that the way that quantum correlations are distributed in a quantum 
register is constrained by the way in which each subsystem gets correlated with the environment.
For a two-qubit system coupled to a common dissipative environment $\mathcal{E}$, we show how to optimise interqubit correlations and entanglement  via a  
quantification of  the qubit-environment information flow, in a process that, perhaps surprisingly, 
does not rely on the knowledge of the state of the environment. 
To illustrate our findings, we consider an optically-driven bipartite interacting qubit $AB$ 
system under the  action of  $\mathcal{E}$. 
By tailoring the light-matter interaction, a relationship between the qubits early stage disentanglement and the qubit-environment entanglement distribution is found. 
We also show that, under suitable initial conditions, the qubits energy asymmetry allows the identification of physical scenarios whereby qubit-qubit entanglement minima coincide with the extrema of the $A\mathcal{E}$ and $B\mathcal{E}$ entanglement oscillations.
}
\\

{\it SUBJECT AREAS: QUANTUM PHYSICS, QUANTUM INFORMATION}

\end{abstract}


\maketitle

The quantum properties of physical systems have been studied  for 
many years as crucial resources for quantum processing tasks and quantum information protocols 
\cite{david,bennett,man,Ladd,Ladd1}. Among these properties, entanglement, non-locality, and correlations 
between quantum objects arise as fundamental features \cite{rev,rev1}. The study of such properties in 
open quantum systems is a crucial aspect of quantum information science~\cite{entropy,entropy1}, in particular because 
decoherence appears as a ubiquituous physical process that prevents the realisation of unitary quantum dynamics---it washes out quantum coherence and multipartite correlation effects, and it has long been recognised as a mechanism responsible for the emergence of classicality from events in a purely quantum realm~\cite{zurekdeco}.  In fact, it is the influence 
of  harmful errors caused by the interaction of a quantum register with its environment 
\cite{zurekdeco,jhdeco,jhdeco1,unruh} that precludes 
the construction of an efficient scalable quantum computer \cite{preskill,preskill1}.

Many works devoted to the study of entanglement and correlations dynamics in open quantum systems 
are focused on the analysis  of the reduced system of interest (the register) and the quantum state of the 
environment is usually  discarded \cite{open,open2,open3,Yu,Yu1}. 
There have  recently been proposed, however, some ideas for detecting system-environment 
correlations (see e.g.,~\cite{gess,modi} and references therein). For example, experimental tests of 
system-environment correlations detection have been recently carried out by means of single trapped ions~\cite{gess2}. The role 
and effect of the system-environment correlations on the dynamics of open quantum systems 
have also been studied within the spin-boson model~\cite{gong,gong2}, and as a precursor of a 
non-Markovian dynamics~\cite{mazz}.
Here, we approach the  qubit-environment dynamics from 
a different perspective and  show that valuable information about the evolution  of quantum entanglement 
and correlations can be obtained if the flow of information  between the register and the environment is better understood.

It is a  known fact that a quantum system composed by many parts cannot freely share entanglement or 
quantum correlations between its parts \cite{fan2,monod,monod1,monod2,monod3,monod4}. Indeed, there are strong constrains on how these 
correlations can be shared, which gives rise to what is known as monogamy of quantum correlations. In 
this paper we use monogamic relations to demonstrate that the way 
that quantum correlations are distributed in a quantum register is constrained by the way in which each 
subsystem gets correlated with the reservoir \cite{mono,KW,fan2}, and that  an optimisation of the  interqubit entanglement and correlations can be devised via a quantification of the  information flow between each 
qubit and its environment.


We consider a bipartite $AB$ system (the qubits) interacting with  a third subsystem $\mathcal{E}$ (the environment). 
We begin by assuming that the {\it whole} `$AB\mathcal{E}$ system' is described  by
an initial pure state  $\rho_{AB\mathcal{E}}(0)=\rho_{AB}(0)\otimes\rho_{\mathcal{E}}(0)$; i.e., at
$t=0$, the qubits  and the environment
density matrices, $\rho_{AB}$ and $\rho_{\mathcal{E}}$, need to be pure. 

The global $AB\mathcal{E}$ evolution is given by
\begin{eqnarray}
\rho_{AB\mathcal{E}}(t) & = & e^{-\frac{i}{\hbar}Ht}\rho_{AB\mathcal{E}}(0)e^{\frac{i}{\hbar}Ht},
\end{eqnarray}
where $H$ denotes the Hamiltonian of the tripartite system.
Since $\rho_{AB\mathcal{E}}(0)$ is pure, $\rho_{AB\mathcal{E}}(t)$ is also pure for all time 
and hence  we can calculate the way that
$AB$ gets entangled with the environment directly from the
entropy. For the entanglement of formation $E_{(AB)\mathcal{E}}$, for example,
this is given by the von Neumann entropy 
$E_{(AB)\mathcal{E}}=S(\rho_{AB})=S(\rho_{\mathcal{E}})$~\cite{Divi,man}.  In order  to quantify 
the way in which $A$ ($B$) gets entangled with $\mathcal{E}$, we calculate $E_{A\mathcal{E}}$ ($E_{B\mathcal{E}}$) 
by means of the Koashi-Winter (KW) relations (see the Methods section)~\cite{KW,fan2} 
\begin{equation}
E_{A\mathcal{E}}  =  \delta_{AB}^{\leftarrow}+S_{A|B},\; E_{B\mathcal{E}}  =  \delta_{BA}^{\leftarrow}+S_{B|A},
\label{entE}
\end{equation}
where $\delta_{ij}^{\leftarrow}$ denotes the quantum discord \cite{zurek,hend,datta}, and $S_{i|j}$ is the 
conditional entropy \cite{rev,rev1}.
Since the tripartite state $AB\mathcal{E}$ remains pure for all time $t$, we can calculate, even without
any knowledge about $\mathcal{E}$, 
the entanglement $E_{ij}$ between each subsystem and the environment. We do so by means
of discord. We also compute the quantum discord between each 
subsystem and the environment as (see the Methods section)
\begin{equation}
\delta_{A\mathcal{E}}^{\leftarrow} = E_{AB}+S_{A|B},\; \delta_{B\mathcal{E}}^{\leftarrow} = E_{AB}+S_{B|A}.
\label{discE}
\end{equation}
We note that, in general, $\delta_{A\mathcal{E}}^{\leftarrow}\neq\delta_{\mathcal{E}A}^{\leftarrow}$ and 
$\delta_{B\mathcal{E}}^{\leftarrow}\neq\delta_{\mathcal{E}B}^{\leftarrow}$, i.e., these 
quantities are not symmetric. Directly from the KW relations, such asymmetry can be understood due to 
the different behaviour exhibited by the entanglement of formation 
and the discord for the $AB$ partition; e.g., $\delta_{\mathcal{E}A}^{\leftarrow}=\delta_{BA}^{\leftarrow}+S_{A|B}$, 
such that $\delta_{A\mathcal{E}}^{\leftarrow}\neq\delta_{\mathcal{E}A}^{\leftarrow}$ when $\delta_{BA}^{\leftarrow}\neq E_{AB}$ 
(the equality holds for bipartite pure states).
In our setup the $AB$ partition goes into a mixed state   due to the dissipative effects and the qubits 
detuning~\cite{jh3,jh4}. In our calculations, the behaviour of $\delta_{i\mathcal{E}}^{\leftarrow}$ and $\delta_{\mathcal{E}i}^{\leftarrow}$, $i=A,B$, 
is  similar, so we only compute, without loss of generality, those correlations given by 
equations~\eqref{discE}. 
An important aspect to be emphasised on the KW relations concerns its definition in terms of the entanglement of formation. Although the original version of the KW relations is given in terms of the entanglement of formation and classical correlations defined in terms of the von Neumann entropy, this is not a necessary condition. Indeed, similar monogamic relations can be determined by any concave measure of entanglement. In this sense, we can define a KW relation in terms of the tangle or even the concurrence, since they both obey the concave property. For instance, in~\cite{osborne} the 
authors use the KW relation in terms of the  linear entropy to show that the tangle is monogamous for a system of $N$ qubits. Here we use the entanglement of formation and quantum discord given their nice operational interpretations, but we stress that this is not a necessary condition.

\begin{figure*}[ht]
  \centering
  \includegraphics[width=17cm]{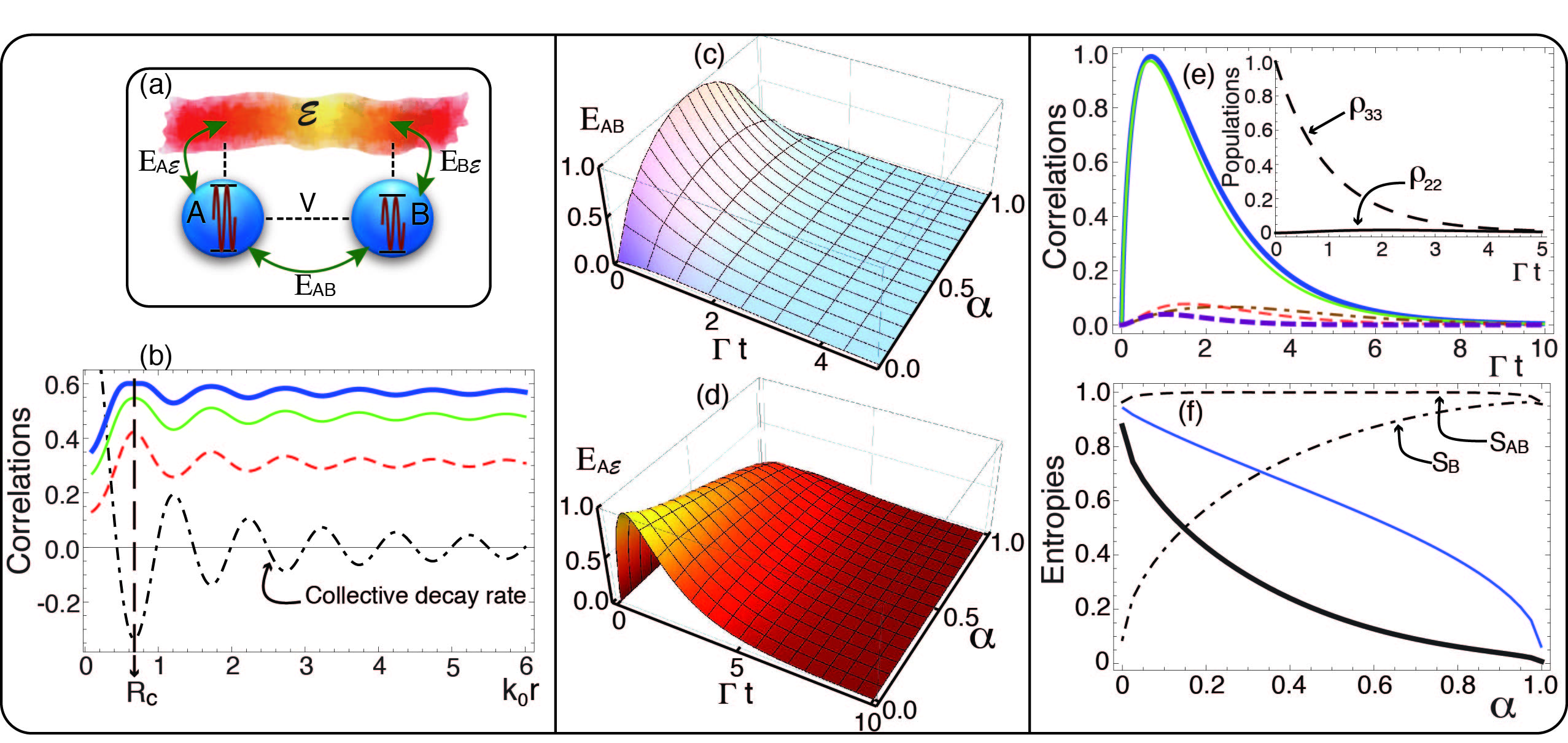}
  \caption{\textbf{Physical setup, quantum entanglement and correlations in the interacting qubit system.} 
  (a) Schematics of the considered $AB\mathcal{E}$ physical system and the flow of quantum information: 
  two two-level emitters (qubits) interacting with a dissipative environment, which are  allowed to be optically-driven via external laser excitation. 
  (b) Quantum discord $\delta^{\leftarrow}_{AB}$ (dashed-red line), $\delta^{\leftarrow}_{A\mathcal{E}}$
  (solid-green line), 
 and entanglement of formation $E_{A\mathcal{E}}$ (solid-blue line); initial state $\ket{\Psi^+}$, at $t =\Gamma ^{-1}$.
  The dotted-dashed curve shows the variation in the collective decay rate $\gamma$ due to changes in the interqubit separation $r$. 
  The vertical line signals the optimal $r=R_c\simeq 0.674\, k_0^{-1}$. 
  (c) Qubit-qubit ($E_{AB}$), and (d) qubit-environment ($E_{A\mathcal{E}}$) entanglement dynamics for the $\alpha$  initial states.
  (e) Quantum discord $\delta^{\leftarrow}_{AB}$
  (dashed-red curve), $\delta^{\leftarrow}_{A\mathcal{E}}$ (solid-green curve), and  $\delta^{\leftarrow}_{B\mathcal{E}}$ (dotted-dashed-brown curve), and entanglement $E_{AB}$ (dashed-purple curve), and $E_{A\mathcal{E}}$ (solid-blue curve).  Populations 
  $\rho_{ii}$ (inset), $\alpha = 0$.  (f) Conditional entropy $S_{A|B}$ 
  (solid-black curve), and $S_{M^{B}_{i}}$ (solid-blue curve) for qubit initial states $\alpha$, $t=\Gamma^{-1}$. $r= R_c$.
} 
  \label{fig1}
\end{figure*}

We illustrate the above statements by considering qubits that are represented by two-level quantum emitters, where  
$\ket{0_{i}}$, and $\ket{1_{i}}$  denote the ground and excited state of emitter $i$, respectively, with 
individual transition frequencies $\omega_{i}$, and in
interaction with a common environment ($\mathcal{E}$) comprised by the vacuum quantised radiation field ~\cite{jh3,jh4}, 
as schematically shown in Fig. \ref{fig1}(a), where $V$ denotes the strength of the interaction between the qubits.

The total Hamiltonian describing the dynamics of the whole $AB\mathcal{E}$ system can be written as
\begin{eqnarray}
H & = & H_{Q}+H_{\mathcal{E}}+H_{Q\mathcal{E}},
\end{eqnarray}
where the qubits free energy $H_Q=-\frac{\hbar}{2}\left(\omega_1\sigma^{(1)}_{z}+\omega_2\sigma^{(2)}_{z}\right)$, the environment 
Hamiltonian $H_{\mathcal{E}}=\sum_{\vec{k}s}\hbar\omega_{\vec{k}s}( \hat{a}^{\dagger}_{\vec{k}s}\hat{a}_{\vec{k}s} + 1/2)$, 
and the qubit-environment interaction, in the dipole approximation,
$H_{Q\mathcal{E}}=-\text{i}\hbar\sum^{2}_{i=1}\sum_{\vec{k}s}\left(\bm{\mu}_{i}~\cdot~\mathbf{u}_{\vec{k}s}(\vec{r}_i)\sigma^{(i)}_{+}\hat{a}_{\vec{k}s}
  + \bm{\mu}^*_{i}\cdot \mathbf{u}_{\vec{k}s}(\vec{r}_i)\sigma^{(i)}_{-}\hat{a}_{\vec{k}s} - \mathrm{H.c.}\right),$
where 
$\sigma^{(i)}_{+}:=\ket{1_{i}}\bra{0_{i}}$, and
$\sigma^{(i)}_{-}:=\ket{0_{i}}\bra{1_{i}}$ are the raising and lowering
Pauli operators acting on the qubit $i$,
$\mathbf{u}_{\vec{k}s}(\vec{r}_i)~=~(\omega_{\vec{k}s}/2\epsilon_{0}\hbar\vartheta)^{1/2}\bm{\varepsilon}_{\vec{k}s}\exp \text{i}\vec{k}\cdot\vec{r_i}$ 
is the coupling constant, $\epsilon_{0}$ the vacuum permittivity, $\vartheta$ 
the quantisation volume, $\bm{\varepsilon}_{\vec{k}s}$ the unitary vector 
of the field mode, $\hat{a}_{\vec{k}s}$ ($\hat{a}^{\dagger}_{\vec{k}s}$) are
the annihilation (creation) operators of the mode, and $\omega_{\vec{k}s}$ 
is its frequency.

For the sake of completeness, we also allow for an external qubit control whereby the qubits can be  optically-driven by a coherent 
laser field of frequency $\omega_{L}$, ${H}_L=\hbar \ell^{(i)}(\sigma^{(i)}_{-}e^{i\omega_{L}t}+\sigma^{(i)}_{+}e^{-i\omega_{L}t})$, 
where  $\hbar\ell_{i}=-\boldsymbol{\mu}_i\cdot \boldsymbol{E}_i$ gives the 
qubit-field coupling, with $\boldsymbol{\mu}_i$ being the $i$-th  transition dipole moment  and 
$\boldsymbol{E}_i$ the amplitude of the coherent driving acting on qubit $i$ located at position $\vec{r}_i$. 
The two emitters are separated by the vector $\vec{r}$ and
are characterised by transition dipole moments $
\bm{\mu}_{i}\equiv\bra{0_i}\mathbf{D}_{i}\ket{1_i}$, with dipole
operators $\mathbf{D}_{i}$, and spontaneous emission rates $\Gamma_i$.

Given the features of the considered physical system, we may assume a weak system-environment coupling such that the  Born-Markov approximation is valid, and we work within the rotating wave approximation for both the 
system-environment and the system-external laser Hamiltonians~\cite{ficek}.  Within this framework, the effective Hamiltonian of the reduced two-qubit $AB$ system, 
which takes into account both the effects of the 
interaction with the environment and the interaction with the coherent laser field, can be written as
\begin{eqnarray}
H_{0} = H_{Q}+H_{12}+H_{L},
\end{eqnarray}
where ${H}_{12} = \half \hbar V \big( 
	\sigma^{(1)}_{x}\otimes\sigma^{(2)}_{x} 
	+ \sigma^{(1)}_{y}\otimes\sigma^{(2)}_{y}\big)$, and $V$ is the strength of the dipole-dipole (qubit) coupling which
depends on the separation and orientation between the dipoles \cite{jh3,jh4,ficek}.
 
In order to impose the pure initial condition to the $AB\mathcal{E}$ system required to use the KW relations, we suppose that the quantum register is in a pure initial state and that we have a zero temperature environment. Thus,
\begin{eqnarray}
\rho_{AB\mathcal{E}}(0)  =  \rho_{AB}\otimes|0\rangle_\mathcal{E}\langle0|.
\end{eqnarray}
However, we note that a less controllable and different initial state for the environment can be 
considered since an appropriate purification of 
the environment $\mathcal{E}$ with a new subsystem $\mathcal{E}'$ could be realised. Despite this, for the sake of simplicity in calculating the quantum register dynamics, we consider a zero temperature environment.

The results below reported require a quantification of the qubits dissipative dynamics. This is described by means of the quantum master equation~\cite{jh3,jh4}:
\begin{equation}
  \label{master}
  {\dot\rho}= C_U- 
  \sum_{i,j=1}^{2}\frac{\Gamma _{ij}}{2} \Big( 
    {\rho} \sigma^{(i)}_{+} \sigma^{(j)}_{-} + 
    \sigma^{(i)}_{+}\sigma^{(j)}_{-}{\rho}
    -2\sigma^{(j)}_{-}{\rho} \sigma^{(i)}_{+} 
  \Big),
\end{equation}
where the commutator $C_U\equiv - \frac{i}{\hbar}  [ {H_0}, {\rho} ]$ gives the unitary part of the evolution. 
The individual and collective spontaneous emission rates are considered such that 
$\Gamma_{ii}=\Gamma_{i}\equiv \Gamma$, and
$\Gamma_{ij}=\Gamma^{\ast}_{ji}\equiv \gamma$, respectively. For simplicity of writing, we adopt the notation $\rho_{ij}$, where $i,j=1,2,3,4$ for the $16$ density matrix elements; $\sum_i\rho_{ii}=1$.

The master equation \eqref{master} gives a solution for $\rho_{AB}(t)$
that becomes mixed since it creates quantum correlations with the environment.
We pose the following questions:  i) How does each qubit  get
entangled with the environment? ii) How does this depend on the energy mismatch 
between $A$ and $B$?, and iii) on the external laser pumping? \\

\noindent
{\bf\large Results}\\
\noindent
{\bf Quantum register-environment correlations.}
To begin with the quantum dynamics of the qubit-environment correlations, we initially consider 
resonant qubits, $\omega_{1}=\omega_2\equiv\omega_0$. 
In the absence of optical driving, there is an optimal inter-emitter 
separation $R_c$ which maximises the correlations~\cite{fan1}. In Fig.~\ref{fig1}(b)  we plot the  quantum discord
$\delta^{\leftarrow}_{A\mathcal{E}}$, and $\delta^{\leftarrow}_{AB}$, and the entanglement of formation $E_{A\mathcal{E}}$ 
as a function of the interqubit separation $k_0 r$ at  $t =\Gamma^{-1}$.
The maximum value reached by each correlation is due to the behaviour of the collective damping $\gamma$, which 
reaches its maximum negative value at the optimal  separation $k_0 R_{c}\simeq 0.674$, as shown in Fig.~\ref{fig1}(b), with 
$k_0=\omega_0/c$. This is due to the fact that the initial 
state $\ket{\Psi^+}=\frac{1}{\sqrt{2}}(\ket{01}+\ket{10})$ decays at the rate $\Gamma +\gamma$ (see equations \eqref{ASolution} with 
$\alpha=1/2$), and hence  the maximum life-time of $\ket{\Psi^+}$
is obtained for  the most negative value of $\gamma$: for any time $t$, the correlations reach their maxima precisely at the 
interqubit distance $R_{c}$ (the same result holds for the $B\mathcal{E}$ bipartition, not shown).

We stress that  it is the collective damping and {\it not} the dipolar 
interaction that defines the distance $R_{c}$.  For a certain family of initial states (which includes 
$\ket{\Psi^+}$), the free evolution of the emitters is independent of the interqubit interaction $V$~\cite{jh2011}: for  the 
initial states $\ket{\Psi(\alpha)}=\sqrt{\alpha}\ket{01}+\sqrt{1-\alpha}\ket{10}$, $\alpha\in [0,1]$, equation~\eqref{master}  
admits an analytical solution and the non-trivial density matrix elements read
\begin{eqnarray}
\rho_{11}(t) &=& 1-\rho_{22}^+(t) -\rho_{33}^-(t) ,\label{ASolution}\\ \nonumber
\rho_{23}(t)&=&\frac{e^{-\Gamma t}}{2}\big[2\sqrt{\alpha(1-\alpha)}\cosh{(\gamma t)} - \sinh{(\gamma t)}\\ 
&&+ \mathrm{i} (2\alpha-1) \sin{(2V t)}\big] , \nonumber  \\  \nonumber 
\left [
\begin{array}{c}
\rho_{22}^+(t)   \\
\rho_{33}^-(t) \\
\end{array}
\right ]
 &=& \frac{e^{-\Gamma t}}{2}\big[\pm  (2\alpha-1) \cos{(2Vt)}+\cosh{(\gamma t)} \\
 && -
2\sqrt{\alpha(1-\alpha)} \sinh{(\gamma t) }\big], \nonumber
\end{eqnarray}
and 
$\rho_{32}(t)=(\rho_{23}(t))^*$. This solution implies that the density matrix dynamics dependence on $V$ 
vanishes for $\alpha = 1/2$ ($\ket{\Psi^+}$), and hence  the damping 
$\gamma$ becomes the only collective parameter responsible for the oscillatory behaviour of the 
correlations, as shown  in Fig.~\ref{fig1}(b). A similar analysis can 
be derived for the initial states $\ket{\Phi(\beta)}=\sqrt{\beta}\ket{00}+\sqrt{1-\beta}\ket{11}$. Thus, the `detrimental' 
behaviour of the system's correlations $\delta_{AB}^{\leftarrow}$ and 
$E_{AB}$ reported in \cite{fan1} is actually  explained 
because such $\beta$ states are not, in general, `naturally' supported by the system's Hamiltonian since 
they are not eigenstates of $H_S=H_Q+H_{12}$.
\begin{figure*}[ht]
  \centering
  \includegraphics[width=12cm]{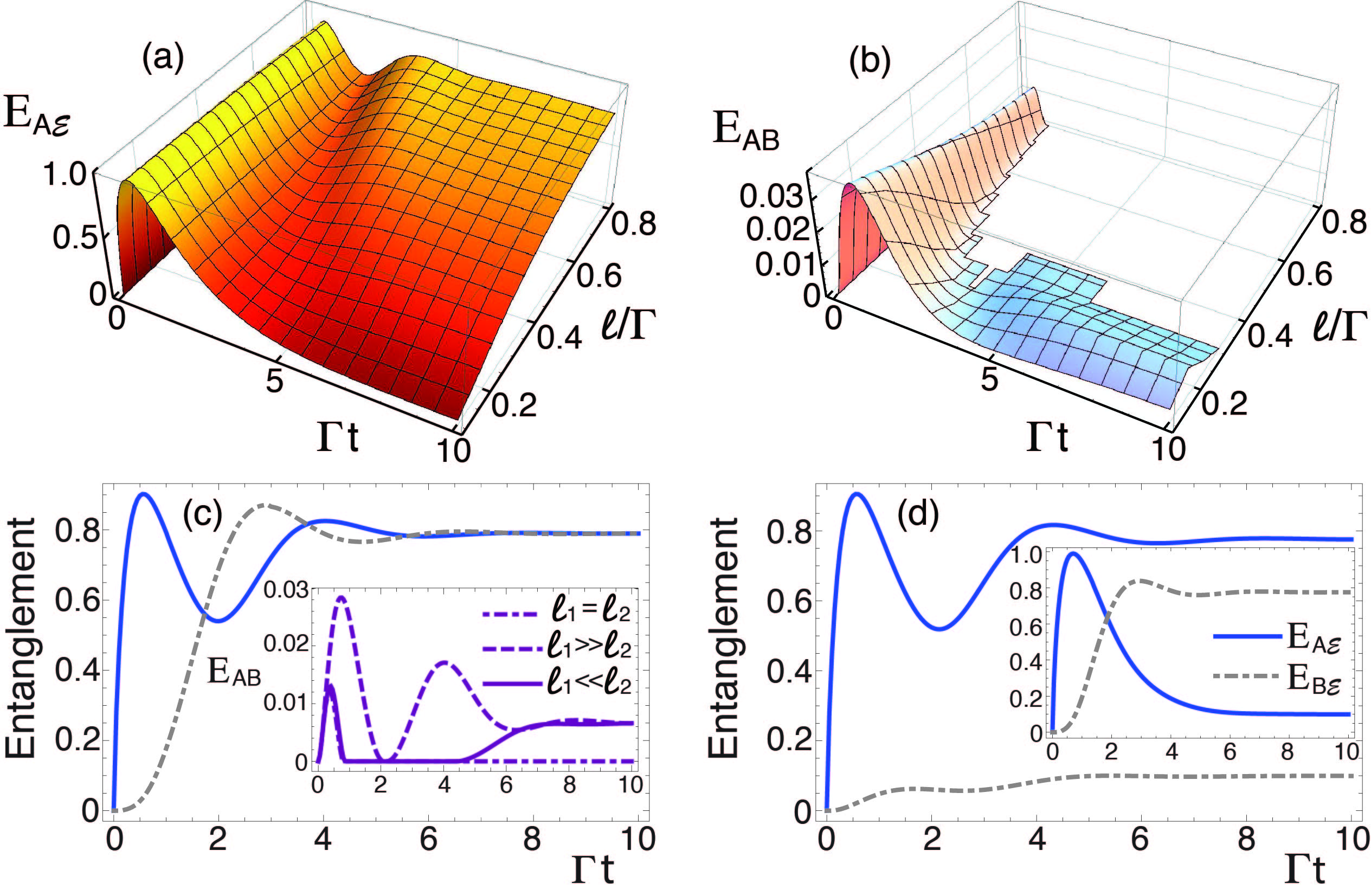}
  \caption{{\bf Driven quantum correlations}. Dynamics of quantum entanglement (a) $A\mathcal{E}$ and  (b) $AB$ as functions of the 
  laser intensity $\ell$. Main (c) $A\mathcal{E}$ (solid) and $B\mathcal{E}$ (dotted-dashed), 
  $\ell_1\equiv\ell_2=\ell=0.8\Gamma$; the inset plots  $E_{AB}$ for three scenarios, main  (d) 
  $\ell_1=0.8\Gamma$, $\ell_2=0$; the inset plots $\ell_2=0.8\Gamma$, $\ell_1=0$. $\rho_{33}(0)=1$, and $r=R_c$. 
} 
  \label{fig5}
\end{figure*}

We now consider the qubits full time evolution and 
calculate the correlations dynamics for the whole spectrum of  initial states $\ket{\Psi(\alpha)}$, $0\leq\alpha\leq1$. 
The emitters' entanglement $E_{AB}$ exhibit an 
asymptotic decay for all $\alpha$ values, with the exception of the 
two limits $\alpha\rightarrow0$ and $\alpha\rightarrow1$, for which the subsystems begin 
to correlate with each other and  the entanglement increases until it reaches a 
maximum before  decaying monotonically, as shown in Fig.~\ref{fig1}(c). The $AB$ discord also follows 
a similar behaviour; this can be seen in Fig.~\ref{fig1}(e) for $\alpha=0$. 
Initially, at $t=0$, the entanglement $E_{A\mathcal{E}}$
(Fig.~\ref{fig1}(d)) equals zero because of the separability 
of the tripartite $AB\mathcal{E}$ state at such time. 
After this, 
the entanglement  between $A$ and $\mathcal{E}$ increases to its maximum, which is reached at a  different time 
($t\sim\Gamma^{-1}$) for each $\alpha$, and then decreases 
asymptotically. The simulations shown in Figs.~\ref{fig1}(c) and (d) have been performed 
for the optimal inter-emitter separation 
$R_{c}$. 
These  allow to access the dynamical qubit  information (entanglement and correlations) exchange between the 
environment and each subsystem for  suitable qubit initialisation.

Figure~\ref{fig1}(c) shows the quantum entanglement between identical emitters: $E_{AB}$
is symmetric with respect to the initialisation $\alpha = 1/2$, i.e.,
the behaviour of $E_{AB}$ is the same for the separable states $\ket{01}$ and 
$\ket{10}$. In contrast, Fig.~\ref{fig1}(d) exhibits a somewhat different behaviour for the  entanglement 
$A\mathcal{E}$, which  is not symmetric with respect to $\alpha$: the maximum reached by $E_{A\mathcal{E}}$ 
increases as $\alpha$ tends to $0$ (the discord $\delta^{\leftarrow}_{A\mathcal{E}}$ follows the same behaviour--not shown). 
The dynamical  distribution of entanglement between the 
subsystem $A$ and the environment $\mathcal{E}$  leads to the following: it is possible to have near zero interqubit  
entanglement (e.g., for the $\alpha = 1$ initialisation)  whilst the entanglement between one subsystem and the 
environment also remains very close to zero throughout  the evolution. 

This result stresses the sensitivity of the qubit-environment entanglement (and correlations) distribution to its qubit initialisation. 
To understand why this is so (cf. states $\ket{01}$ and $\ket{10}$),  
we analyse the expressions for 
$\delta^{\leftarrow}_{A\mathcal{E}}$ 
and $E_{A\mathcal{E}}$. From equations~\eqref{entE} and equations~\eqref{discE}, and since $E_{AB}$ and $\delta^{\leftarrow}_{AB}$ are both 
symmetric, the asymmetry of $\delta^{\leftarrow}_{A\mathcal{E}}$ and $E_{A\mathcal{E}}$ should follow from  the conditional entropy 
$S_{A|B}=S(\rho_{AB})-S(\rho_{B})$. This is plotted in the solid-thick-black curve in Fig.~\ref{fig1}(f). The  behaviour of the 
conditional entropy is thus reflected in the dynamics 
of quantum correlations and entanglement  between 
$A$ and 
$\mathcal{E}$, and this can be seen if we compare the behaviour of $E_{A\mathcal{E}}^{\leftarrow}$ around $t=\Gamma^{-1}$ 
throughout the $\alpha$-axis in Fig.~\ref{fig1}(d), with that of $S_{A|B}$ shown in 
Fig~\ref{fig1}(f). Since this conditional entropy gives the amount of partial information 
that needs to be transferred 
from $A$ to $B$ in order to  know $\rho_{AB}$, with a prior 
knowledge of the state of $B$ \cite{HOW}, we have shown that this amount of 
information may be extracted from the dynamics of the quantum correlations generated 
between the qubits and their environment.

Interestingly,  by replacing the definition of 
$\delta^{\leftarrow}_{AB}$ \cite{zurek} into the first equality of equations~\eqref{entE}, we find that the entanglement 
of the $A\mathcal{E}$ partition is exactly the post-measure conditional entropy of the $AB$ partition: 
\begin{equation}
\label{AEcone}
E_{A\mathcal{E}}=S_{M^{B}_{i}}\equiv{\rm min}_{M^{B}_{i}}\Big[\sum_{i}p_{i}S\left(\rho_{A|i}\right)\Big] ,
\end{equation}
that is, the entanglement between the emitter $A$ and its environment is the conditional 
entropy of  $A$ after 
the partition  $B$ has been measured, and hence the asymmetric behaviour of $E_{A\mathcal{E}}$ can be 
verified by plotting this quantity, as shown by the solid-blue curve of Fig.~\ref{fig1}(f).
A physical reasoning for the asymmetric behaviour of the $A\mathcal{E}$ 
correlations points out that for $\alpha \rightarrow 0$ the state $\ket{10}$ 
has higher weights throughout the whole dynamics. For instance, for $\alpha=0$ the 
subsystem $B$ always remains close to its ground state, and 
transitions between populations $\rho_{22}$ and $\rho_{33}$ do not take place, as it is shown   
in the inset of Fig.~\ref{fig1}(e). This  means that partition $B$ keeps almost inactive during
this specific evolution and therefore does not share much information, neither 
quantum nor locally accessible with partition $A$ and the environment $\mathcal{E}$. This can be seen  from  
the quantum discord $\delta^{\leftarrow}_{B\mathcal{E}}$, which is plotted as the dotted-dashed-brown curve of 
Fig.~\ref{fig1}(e). We stress that this scenario allows $A$ to 
get strongly correlated with the environment $\mathcal{E}$.
\begin{figure}[t]
  \centering
  \includegraphics[width=\columnwidth]{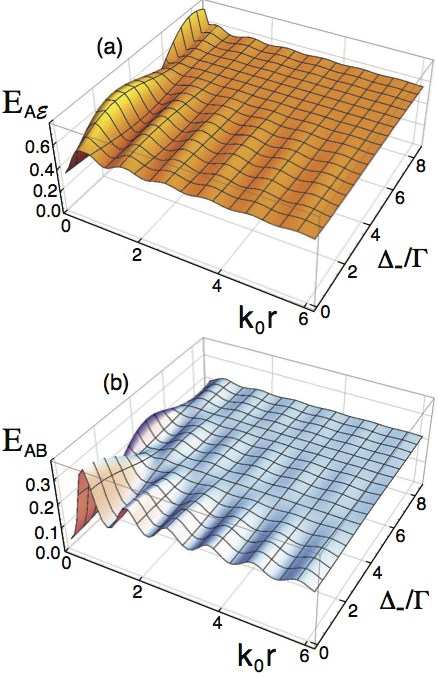}
  \caption{{\bf Quantum entanglement for detuned qubits.} (a)  $E_{A\mathcal{E}}$, and (b) $E_{AB}$ dependence on the inter-emitter distance $r$ for frequency detuning $\Delta_-\equiv \omega_1-\omega_2$,  and  qubits
  initial state $\ket{\Psi^+}$, $t=\Gamma^{-1}$. 
} 
  \label{fig6}
\end{figure}

Although $E_{A\mathcal{E}}$ and $\delta^{\leftarrow}_{A\mathcal{E}}$ are 
not `symmetric' with  respect to 
$\alpha$,  it is the information flow, i.e., the way the information gets transferred 
between the qubits and  the environment, the quantity 
that recovers the symmetry exhibited  by $E_{AB}$ 
in Fig.~\ref{fig1}(c). In other words, 
if the initial state were $\ket{01}$, or in general, $\alpha\rightarrow1$, the partition 
$A$ would remain almost completely inactive and  the flow of information 
would arise from the bipartite partition $B\mathcal{E}$ instead of $A\mathcal{E}$.  A simple numerical 
computation for $\alpha=0$ at $t=\Gamma ^{-1}$ shows that
\begin{eqnarray}
	\rho_A=
	\left(
		\begin{array}{cc}
			0.62 & 0 \\
			0 & 0.38
		\end{array}
	\right) 
	\;\; \mathrm{and} \;\;\; 
	\rho_B=
	\left(
		\begin{array}{cc}
			0.99 & 0 \\
			0 & 0.01
		\end{array}
	\right) ,
\end{eqnarray}
with $S_A=0.96$ and $S_B=0.09$, respectively. This means that the state of subsystem $B$ is close to 
a pure state (its ground state), and no much information about it may be gained.  Instead,  almost all 
the partial information on the state of $A$ can be caught regardless of whether the system $B$ is measured or not. 
From this simple reasoning, and by means of the KW relations, the results shown in 
Figs.~\ref{fig1}(c-f) arise. The opposite feature between $\rho_A$ and $\rho_B$ occurs for  $\alpha=1$, and  in this case, it is the partition $B\mathcal{E}$ that plays the 
strongest correlation role.\\

\noindent
{\bf Information flow in laser-driven resonant qubits.}
$H_L$ conveys an additional degree of control of the qubits information (entanglement and discord) flow. 
Let us consider a continuous laser field acting 
with the same amplitude, $\ell_1=\ell_2\equiv\ell$, on the two emitters, and in resonance with the 
emitters' transition energy, $\omega_L = \omega_0$. The subsystem $A$ gets the strongest 
correlated with the 
environment for the initial pure  state $\ket{10}$ in the 
relevant  time regime (see Fig.~\ref{fig1}(d)), but this correlation monotonically decays  to zero in the steady-state 
regime. In Fig.~\ref{fig5} we see the effect of the laser driving for the  initial state  $\ket{10}$, for qubits separated 
by the optimal distance $r=R_c$. The laser field removes the monotonicity  in the entanglement and correlations 
decay between $A$ and $\mathcal{E}$, and, as shown in Fig.~\ref{fig5}(a), the more intense the 
laser radiation (even at the weak range $\ell\leq\Gamma$), the 
more entangled the composite $A\mathcal{E}$ partition becomes. This translates, in turn, into a dynamical  mechanism 
in which the qubit register $AB$ gets  rapidly disentangled and,  even at couplings as  weak as  $\ell\sim 0.4\Gamma$, 
the qubits exhibit
early stage disentanglement, as shown in Fig. \ref{fig5}(b). This regime coincides with the appearance of oscillations in 
the $A\mathcal{E}$ entanglement (see Fig.~\ref{fig5}(a)), and steady nonzero $A\mathcal{E}$ entanglement translates 
into induced interqubit ($AB$) entanglement  suppression by means of the laser field. 

By tailoring  the laser amplitude we are able to induce and control the way in which the qubits get correlated with each 
other and with the environment.~The graphs~\ref{fig5}(c) and (d) show three different scenarios in terms of such amplitude. 
In graph  (c) we plot $E_{A\mathcal{E}}$ (solid-blue curve) and $E_{B\mathcal{E}}$ (dashed-dotted-grey curve) for the 
symmetric light-matter interaction ($\ell_1=\ell_2\equiv\ell=0.8\Gamma$), which leads to  ESD in the partition $AB$ 
(see graph (b)), as well as to a symmetric qubit-environment correlation in the stationary regime. However, as can be 
seen in main  graph (d), where we have assumed $\ell_1>>\ell_2$, the breaking of this symmetry completely modifies 
the qubit-environment entanglement, and now  it is qubit $A$ that  gets strongly correlated with the environment, while 
qubit $B$ remains weakly correlated during the dynamics. The opposite arises for $\ell_1<<\ell_2$ (inset of panel (d)): 
$E_{B\mathcal{E}}$  becomes much higher than $E_{A\mathcal{E}}$, which decays monotonically after reaching its 
maximum. Remarkably, we notice that these two asymmetric cases lead a nonzero qubit-qubit entanglement as shown 
in the inset of graph (c), where equal steady entanglement is obtained. 
It means that the qubits early stage disentanglement \cite{Yu,Yu1} can be interpreted in terms of the 
entanglement distribution between the qubits and the environment. 
We interpret this behaviour  as the flow and 
distribution of entanglement in the different partitions of the whole tripartite system \cite{fanchini2}, and hence this result 
shows that an applied external field  may be used  to dictate the flow of quantum information within the full tripartite system.\\

\noindent
{\bf Flow of information in detuned qubits.}
We now consider a more general scenario in which each  
two-level emitter is resonant at a different transition energy, and hence a 
molecular detuning $\Delta_-=\omega_1-\omega_2$ arises; $\omega_0\equiv (\omega_1+\omega_2)/2$.
Such a detuning substantially modifies the qubit-qubit and qubit-environment correlations. Since $\Delta_-\neq0$, 
the $\alpha=1/2$-time independence 
of equations~\eqref{ASolution} with respect to $V$ no longer holds, and the  critical distance $R_c$ of Fig.~\ref{fig1}(b) becomes  strongly modified: 
the information flow exhibits a more involved dynamics precisely at distances  
$r<R_c$, and 
the intermediate sub- and super-radiant states are no longer the maximally entangled Bell  states.

As shown in Fig. \ref{fig6}, the 
oscillations of $A\mathcal{E}$ and $AB$ entanglement (and their maxima)
start to decrease and become flat as the molecular detuning rises ($\Delta_-=0$ corresponds to the case shown in Fig.~\ref{fig1}(b)). This 
means that now, it is not only the collective decay rate $\gamma$ that modulates 
the behaviour of the entanglement and the correlations, but also the interplay between the detuning  $\Delta_-$ and 
the dipole-dipole interaction $V$. 
Note from Fig.~\ref{fig6} 
that the critical distance $R_c$ for which both the correlations of partition 
$AB$ and those of $A\mathcal{E}$ get their maxima disappears with the inclusion of the 
molecular detuning, and  $E_{AB}$ and $E_{A\mathcal{E}}$ exhibit maxima at different inter-emitter distances as the 
detuning increases: $E_{AB}$ remains global maximum for resonant qubits ($\Delta_-=0$) whereas  $E_{A\mathcal{E}}$ 
reaches its global maximum for a certain $\Delta_-\neq 0$ (e.g., $\Delta_-/\Gamma=8$ at $k_0r\sim0.1$), a value for which 
$E_{AB}$ is  stationary for almost all interqubit separation $r$.

\begin{figure}[t]
  \centering
  \includegraphics[width=\columnwidth]{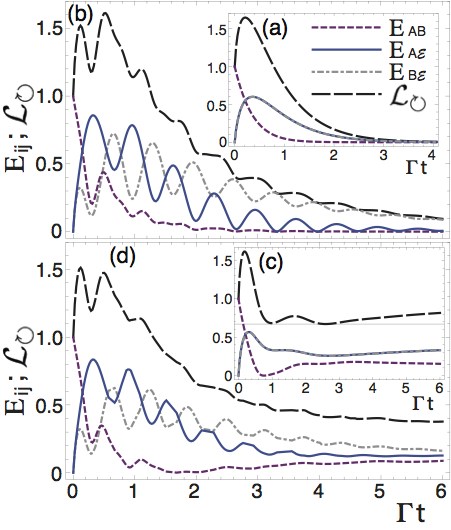}
  \caption{{\bf Quantum correlations and local inaccessible information}. Entanglement dynamics $E_{AB}$, $E_{A\mathcal{E}}$, $E_{B\mathcal{E}}$, 
  and the flow of quantum information $\mathcal{L}_{\circlearrowright}$. 
(a) identical, and (b) detuned emitters, $\Delta_-=8\Gamma$; no laser excitation. 
$\ell=0.8\Gamma$ and  $\omega_0=\omega_L$-laser-driven (c) identical, and  
  (d) detuned emitters, $\Delta_-=8\Gamma$. The initial state $\ket{\Psi^+}$, and $k_0 r=0.1$.}
  \label{fig7}
\end{figure}

To complete the analysis of the general tripartite $AB\mathcal{E}$ system, we now consider that the asymmetric (detuned) qubits are  driven by an 
external laser field on resonance  with the average qubits transition energy, $\omega_0=\omega_L$, 
as shown in Fig.~\ref{fig7} for the two-qubit initial state $\ket{\Psi^+}$. We have plotted the entanglement 
dynamics $E_{AB}$, $E_{A\mathcal{E}}$, and $E_{B\mathcal{E}}$. 
Fig.~\ref{fig7}(a) shows the entanglement evolution for two identical emitters in the absence 
of the external driving. The molecular detuning, and  the laser excitation have been included 
in graphs (b) and (c), respectively. The panel (d) shows the entanglement evolution under detuning and 
laser driving. The monotonic decay of $E_{ij}$ for resonant qubits in Fig.~\ref{fig7}(a) is in clear contrast 
with the $E_{ij}$-oscillatory behaviour due to  the qubit asymmetry, as plotted in Fig.~\ref{fig7}(b). A non-zero 
resonant steady entanglement
is obtained thanks to the continuos laser excitation (Fig.~\ref{fig7}(c) and (d)). 
These graphs have been compared with that of the clockwise flow of pairwise locally inaccessible information 
$\mathcal{L}_{\circlearrowright}=\delta^{\leftarrow}_{B\mathcal{E}}+\delta^{\leftarrow}_{AB}+\delta^{\leftarrow}_{\mathcal{E}A}$ 
\cite{fanchini2}, as shown in the long-dashed black curve.\\

\noindent
{\bf\large Discussion}\\
\noindent
We can now interpret the entanglement dynamics of the $AB$ partition by means of the dynamics of the clockwise
quantum discord distribution in the full tripartite system ($\mathcal{L}_{\circlearrowright}$), and that  of the entanglement of $A\mathcal{E}$ 
and $B\mathcal{E}$ partitions. From the conservation law \cite{fan2} between the distribution of the entanglement of 
formation and discord followed from Eqs. \eqref{entE} and \eqref{discE}, and noting that  
$\mathcal{L}_{\circlearrowright}=\mathcal{L}_{\circlearrowleft}$ for pure states \cite{fanchini2}, where
$\mathcal{L}_{\circlearrowleft}=\delta^{\leftarrow}_{BA}+\delta^{\leftarrow}_{\mathcal{E}B}+\delta^{\leftarrow}_{A\mathcal{E}}$,
a direct connection between qubit-qubit 
entanglement and qubit-environment entanglement can be established \cite{fanchini2}:
\begin{equation}
	E_{AB}=\mathcal{L}_{\circlearrowright}-E_{A\mathcal{E}}-E_{B\mathcal{E}} .
	\label{LII1}
\end{equation}

We note from equation \eqref{LII1} and from the pairwise locally inaccessible information that by
knowing $\delta_{AB}^{\leftarrow}$ (or $\delta_{BA}^{\leftarrow}$), we can exactly compute the qubit-qubit entanglement in terms of the system
bipartitions $A\mathcal{E}$ and $B\mathcal{E}$. 
In particular, we show how a profile of the qubit-qubit entanglement 
might be
identified from the partial information obtained from $E_{A\mathcal{E}}$ and $E_{B\mathcal{E}}$, as indicated in the right-handside of equation~\eqref{LII1}, and shown in Figs.~\ref{fig7}(b) and (d) whereby the local minima of $E_{AB}$ occur at times for which 
the extrema of $E_{A\mathcal{E}}$ and $E_{B\mathcal{E}}$ take place.
However, it is interesting to note that the locally inaccessible information 
$\mathcal{L}_{\circlearrowright}$, which gives a global information 
of the whole tripartite system (the distribution of quantum correlations-discord), can be extracted directly from the quantum state of the register.   This fact can be demonstrated by replacing equation \eqref{AEcone}, and its equivalent formula for the bipartition $B\mathcal{E}$ 
($E_{B\mathcal{E}}=S_{M^{A}_{i}}\equiv{\rm min}_{M^{A}_{i}}\sum_{i}p_{i}S(\rho_{B|i})$), 
into equation \eqref{LII1}:
\begin{equation}
	\mathcal{L}_{\circlearrowright}=E_{AB} + S_{M^{A}_{i}} + S_{M^{B}_{i}}.
	\label{LII2}
\end{equation}
\noindent
This relation means that the entanglement of partition $AB$ plus {\it local accessible} information of subsystems $A$ and $B$, i.e. the post-measured conditional entropies $S_{M^{A}_{i}}$ and $S_{M^{B}_{i}}$, give complete information about the flow of the locally inaccessible 
(quantum) information.

To summarise, we have shown that the way in which quantum systems correlate or share information 
can be understood from the dynamics of the register-environment correlations. This has been done via the KW 
relations established for the entanglement of formation and the quantum discord. 
Particularly, we have shown that the distribution of entanglement between each qubit and the environment signals 
the results for 
both the prior- and post-measure conditional entropy (partial information) shared by the qubits. As a 
consequence of this link, and in particular equation Fig. \eqref{AEcone}, we have also shown that some information 
(the distribution of quantum correlations---$\mathcal{L}_{\circlearrowright}$) about the 
whole tripartite system \cite{fanchini2} can be extracted by performing local operations over one of the 
bipartitions, say $AB$, and by knowing the entanglement of formation in the same subsystem (equation \eqref{LII2}).  We stress 
that these two remarks are completely independent of the considered physical model as they have been deduced from the 
original definition of the monogamy KW relations (see the Methods section). On the other hand, 
considering the properties of the specific model  here investigated (which may be applicable to atoms, small 
molecules, and quantum dots arrays), the study of  the dynamics of the 
distribution of qubit-environment correlations led us to establish that 
qubit energy asymmetry induces entanglement oscillations, 
and that we can extract partial information about $AB$ entanglement by analysing the way in which information 
(entanglement and discord) flows between 
each qubit and the environment, for suitable initial states. 
Particularly, we have shown that the qubits early stage disentanglement may be understood in terms of the laser strength asymmetry which determines the entanglement distribution between the qubits and the environment.  In addition, we have also shown that 
the extrema of the qubit-environment $A\mathcal{E}$ and $B\mathcal{E}$ entanglement oscillations
exactly match the $AB$ entanglement minima.
The study here presented has been done 
without need to explicitly invoque any knowledge about  the state of the environment at any time $t>0$. 

An advantage of using the information gained from the system-environment correlations to get information about the reduced system's entanglement dynamics is that new interpretations 
and understanding of the system dynamics may arise. For instance, one of us~\cite{fanchini3} has used this 
fact to propose an alternative way of detecting the non-Markovianity of an open quantum system by testing 
the accessible information flow between an ancillary system and the local environment of the apparatus (the 
open) system.\\

\noindent
{\bf\large Methods}\\
We give a brief introduction to the monogamy relation between the 
entanglement of formation and the classical correlations established by Koashi and Winter: As a theorem, KW established a trade-off between the entanglement of formation and the 
classical correlations defined by Henderson and Vedral \cite{hend}. They proved that \cite{KW}:

\medskip

\textbf{Theorem} \textit{When $\rho_{AB'}$ is B-complement to $\rho_{AB}$,}

\begin{equation}
	\label{kw1}
	E(\rho_{AB}) + \mathcal{J}^\leftarrow(\rho_{AB'}) = S(\rho_A) ,
\end{equation}
where {\it B-complement} means that there exist a tripartite pure state $\rho_{ABB'}$ such that 
$\mathrm{Tr}_B[\rho_{ABB'}]=\rho_{AB'}$ and $\mathrm{Tr}_{B'}[\rho_{ABB'}]=\rho_{AB}$, where 
$S_{A}:=S(\rho_{A})$ is the von Neumann entropy of the density matrix
$\rho_A\equiv\mathrm{Tr}_B[\rho_{AB}]=\mathrm{Tr}_{B'}[\rho_{AB'}]$, $E_{AB}:=E(\rho_{AB})$ 
is the entanglement of formation, and $\mathcal{J}^\leftarrow_{AB'}:=\mathcal{J}^\leftarrow(\rho_{AB'})$ 
leads the classical correlations.


For our purpose we only show some steps in the proof of the KW relation (equation \eqref{kw1}); the 
complete proof can be straightforwardly followed in \cite{KW}. 
By starting with the definition of the entanglement of formation:
\begin{equation}
	E_{AB}=\min_{\{p_i,\ket{\psi_i}\}}\sum_i p_i S\left(\mathrm{Tr}_B[\proj{\psi_i}]\right) ,
\end{equation}
where the minimum is over the ensamble of pure states $\{p_i,\ket{\psi_i}\}$ satisfying 
$\sum_ip_i\proj{\psi_i}=\rho_{AB}$, it is possible to show, after some algebra, that
\begin{equation}
	\label{sd1}
	\mathcal{J}^{\leftarrow}_{AB'}\geq S_A-\sum_i p_i S\left(\mathrm{Tr}_B[\proj{\psi_i}]\right) .
\end{equation}
Conversely, from the definition of classical correlations \cite{hend}:
\begin{equation}
	\label{ccorr}
	\mathcal{J}^{\leftarrow}_{AB'}=\max_{\{M_k^{B'}\}}\left[S_A-\sum_kp_kS(\rho_{A|k})\right] ,
\end{equation}
where $\rho_{A|k}=\mathrm{Tr}_{B'}\left[\left(\mathbb{1}_A\otimes M_k^{B'}\right)\rho_{AB'}\right]/p_k$ is the 
state of party $A$ after the set of measurements $\{M_k^{B'}\}$ has been done on party $B'$ with 
probability $p_k=\mathrm{Tr}\left[\left(\mathbb{1}_A\otimes M_k^{B'}\right)\rho_{AB'}\right]$. Let us assume 
that $\{M_i^{B'}\}$ is the set achieving the maximum in equation \eqref{ccorr} such that one can write
$\mathcal{J}^{\leftarrow}_{AB'}=S_A-\sum_ip_iS(\rho_{A|i})$, as the 
operators $M_i^{B'}$ may be of rank larger than one, by decomposing them 
into rank-1 nonnegative operators such that 
$M_i^{B'}=\sum_jM_{ij}^{B'}$, one can show the following
\begin{equation}
	\label{sd2}
	E_{AB}\leq\sum_{ij}p_{ij}S\left(\mathrm{Tr}_{B'}[\proj{\phi_{ij}}]\right) ,
\end{equation}
where $\{p_{ij},\ket{\phi_{ij}}\}$ is an ensamble of pure states with $\sum_{ij}p_{ij}\proj{\phi_{ij}}=\rho_{AB'}$ 
as the set of measurements $\{M_{ij}^{B'}\}$ is applied to the pure tripartite state $\rho_{ABB'}$. 
The relationship 
between the sets $\{M_{i}^{B'}\}$ and $\{M_{ij}^{B'}\}$ is through: $p_i=\sum_jp_{ij}$ and 
$p_i\rho_{A|i}=\sum_jp_i\rho_{A|{\{ij\}}}$. 

\medskip

Noting that 
$S_A-\sum_{ij}p_{ij}S(\rho_{A|\{ij\}})\geq S_A-\sum_ip_iS(\rho_{A|i})\equiv\mathcal{J}^{\leftarrow}_{AB'}$ due to the concavity of the von Neumann entropy, but that the opposite inequality arises by the own 
definition of the classical correlations, one concludes that 
$S_A-\sum_{ij}p_{ij}S(\rho_{A|\{ij\}}) = \mathcal{J}^{\leftarrow}_{AB'}$. Then, by putting together the 
results of equations \eqref{sd1} and \eqref{sd2}, one achieves the equation \eqref{kw1}. 

\medskip

By introducing the definition of quantum discord \cite{zurek}:
\begin{equation}
	\delta^{\leftarrow}_{AB'}:=\delta^{\leftarrow}(\rho_{AB'})=\mathcal{I}_{AB'}-\mathcal{J}^{\leftarrow}_{AB'} ,
\end{equation}
where $\mathcal{I}_{AB'}=S_A+S_{B'}-S_{AB'}$ is the quantum mutual information of the bipartition $AB'$, 
into equation \eqref{kw1}, one gets 
\begin{equation}
	\label{kw3}
	E_{AB}-\delta^{\leftarrow}_{AB'}=S_{A|B'} , 
\end{equation}
with $S_{A|B'}=S_{AB'}-S_{B'}$ the conditional entropy. Noting that $S_{A|B'}=-S_{A|B}$ because  the tripartite state 
$\rho_{ABB'}$ is pure, and changing the subscript $B'$ to $\mathcal{E}$, equation  \eqref{kw3} gives rise to the expression 
for $\delta^{\leftarrow}_{A\mathcal{E}}$ in equation \eqref{discE}. The rest of equalities in equations (2) and (3) 
are obtained by moving the three subscripts and applying the corresponding classical correlations to the appropriate 
bipartition.
\\

\noindent
{\bf\large Acknowledgements}\\
\noindent
J.H.R.  gratefully acknowledges 
Universidad del Valle for a leave of absence and for partial funding under grant CI 7930, and the Science, Technology 
and Innovation Fund-General Royalties System (FCTeI-SGR) under contract BPIN 2013000100007. C.E.S. thanks COLCIENCIAS for a fellowship. 
F.F.F. thanks support from the National
Institute for Science and Technology of Quantum Information (INCT-IQ) under  grant  2008/57856-6, the National Counsel of Technological and Scientific Development (CNPq) under 
grant 474592/2013-8, and the S\~ao Paulo Research Foundation (FAPESP) under grant 2012/50464-0.

\end{document}